\documentclass[prd,twocolumn]{revtex4}
\usepackage{graphicx, epsfig}
\usepackage{color}
\usepackage{mathrsfs}
\usepackage{bm}
\usepackage{amsmath,amssymb,yhmath}
\usepackage[caption=false]{subfig}
\usepackage{hyperref}
\usepackage[normalem]{ulem}
\usepackage{mathtools}

\newcommand{\be}{\begin{equation}}
\newcommand{\ee}{\end{equation}}
\newcommand{\ba}{\begin{eqnarray}}
\newcommand{\ea}{\end{eqnarray}}

\newcommand{\bfk}{{\bf k}}

\newcommand{\bfx}{{\bf x}}

\begin{document}
\title{Decay of Cosmic Global String Loops}
\author{Ayush Saurabh$^\dag$, Tanmay Vachaspati$^\dag$, Levon Pogosian$^*$}
\affiliation{
$^\dag$Physics Department, Arizona State University, Tempe, AZ 85287, USA. \\
$^*$Physics Department, Simon Fraser University, Burnaby, BC V5A 1S6, Canada. \\
}

\begin{abstract}
\noindent
We numerically study the decay of cosmic global string loops due to radiation 
of Goldstone bosons and massive scalar ($\chi$) particles. The length of loops
we study range from 200-1000 times the width of the string core. We find that 
the lifetime of a loop is $\approx 1.4 L$.
The energy spectrum of Goldstone boson radiation has a  $k^{-1}$ fall off, where 
$k$ is the wavenumber, and a sharp peak at $k \approx m_\chi/2$,
where $m_\chi$ is the mass of $\chi$.
The latter is a new feature and implies a peak at high energies (MeV-GeV) in 
the cosmological distribution of QCD axions.
\end{abstract}

\maketitle

If high energy particle physics contains a global U(1) symmetry that
spontaneously breaks at lower energies, the universe would be left
with a network of global strings~\cite{Vilenkin:2000jqa}. Loops of such
cosmic global strings would oscillate and decay by the emission of
massless and massive radiation that could form part of the dark
matter density today~\cite{Vilenkin:1986ku}. The scenario is relevant 
to models in which axions are proposed as a means to solve the strong 
CP problem and is frequently 
studied in this context~\cite{Vilenkin:1986ku,Davis:1986xc,Harari:1987ht,
Davis:1989nj,Hagmann:1990mj,Battye:1993jv,Battye:1994au,Yamaguchi:1998gx,
Hagmann:2000ja,Hiramatsu:2010yu, Klaer:2017ond,Klaer:2019fxc}.

The emission of Goldstone radiation from global string loops has 
been investigated in the Nambu-Goto limit in which the string core
has negligible thickness~\cite{Vilenkin:1986ku}. The calculation is enabled by
replacing the scalar field for the Goldstone degree of freedom by
a two form gauge field, leading to the Kalb-Ramond description
of the string~\cite{Kalb:1974yc}. The radiation has also been
studied by numerically evolving the field theory configuration of
a string loop~\cite{Hagmann:1990mj,Hagmann:2000ja}. Both methods 
have their limitations. The
Kalb-Ramond approach assumes that there is only Goldstone
boson radiation and no massive scalar radiation. Also, the backreaction
of Goldstone boson radiation on the string dynamics has not yet
been taken into account.
The numerical field theory method involves a full description
of the problem but is limited in dynamic range. Hence the results
must be extrapolated to long loops such as would arise in cosmology.

Here we adapt our numerical field theory study of local cosmic 
string loops \cite{Matsunami:2019fss} to the case of global strings. 
The key features
of our work that distinguish it from the earlier field theory loop 
simulations ~\cite{Hagmann:1990mj,Hagmann:2000ja}
are that (i) we consider loops that are formed from the collision
of long straight strings, much as they might form in cosmology, and (ii) we
consider relatively long loops, several hundred times the width
of the string core. The first feature helps to set up the initial conditions,
as the straight string solution is known and we only need to patch
together the solutions. The second feature helps with the extrapolation
to cosmological scales.

Our results are summarized as follows. Global string loops emit both
massless Goldstone radiation and massive particles (denoted by $\chi$) 
and decay in a time proportional to the size of the loop $L$. 
Initially our loops have length $4L$ and invariant length (energy divided
by string tension) of 5L due to a Lorentz boost factor of 1.25. 
We find that
the energy distribution of massive particles is peaked at very low
wavenumbers and they are non-relativistic at
production. Eventually the massive particles will decay into 
Goldstone bosons but their decay leaves a sharp feature in the
spectrum of Goldstone bosons.
The energy spectrum of radiated Goldstone bosons takes the form,
\be
\frac{d{\cal E}_k}{dk} = \eta^2 L \frac{a}{k}, \ \ (2L)^{-1} \le k \lesssim  m_\chi
\label{En}
\ee
where, $\eta$ is the vacuum expectation value of the scalar field, 
$k$ is the magnitude of the wavevector, and 
$a \approx 4.8$ is a coefficient that we determine numerically.
Although the spectrum is peaked at the smallest wavevector, the
integrated energy at larger momenta is greater for large loops
because this contribution grows as $\ln (m_\chi L)$.

This paper is organized as follows. In Sec.~\ref{model} we 
describe the field theory, some basic features of cosmic strings,
and our scheme for producing a global string loop. We then
turn to our numerical implementation on a lattice with periodic
boundary conditions (PBC) in Sec.~\ref{simulationresults},
where we also describe the results of our simulations, contrasting 
global string evolution with that of gauge strings.  
In Sec.~\ref{components} we turn to the radiation produced by the
loops and evaluate the fraction of energy in massless to massive
radiation. The spectrum of radiation is discussed in Sec.~\ref{spectrum}.
Here we analyze the spectral features
of both massive and massless radiation from global strings, and find a
good fit to the form of the spectrum in \eqref{En} for the Goldstone
radiation. We conclude in Sec.~\ref{conclusions}, where we also
place our results in the context of earlier work.
The implementation of periodic boundary conditions requires care
as the Goldstone boson cloud around the string loop decays 
relatively slowly and reaches the boundaries of the lattice.
We describe our implementation in Appendix~\ref{pbc}.

\section{Model}
\label{model}

We consider the global U(1) field theory with a complex scalar field, 
$\phi =\phi_1+i\phi_2$, for which the field equations of motion are
\be
\partial_t^2 \phi_a = \nabla^2 \phi_a - \lambda (\phi_b\phi_b- \eta^2)\phi_a 
\label{phieq}
\ee
where $a=1,2$, $\lambda$ is a coupling constants. By suitable
rescalings of the fields and the coordinates, we can set $\lambda=1/2$ 
and $\eta=1$ and then the (classical) model has no free parameters.

The solution for a straight global string along the $z-$axis is
\be
\phi = \eta f(r) e^{i\varphi}, 
\label{stringsoln}
\ee
where $r=\sqrt{x^2+y^2}$, $\varphi=\tan^{-1}(y/x)$, and
$f(r)$ is a string profile functions that vanishes at the origin and 
asymptotes to 1 as
\be
f(r) \to 1 - \mathcal{O} \left ( \frac{1}{r^2} \right ).
\label{profile}
\ee

The energy density in the scalar field is given by
\be
{\cal E} = \frac{1}{2} |\partial_t \phi |^2+\frac{1}{2} |\nabla \phi |^2 
+ \frac{\lambda}{4} (|\phi|^2-\eta^2)^2
\ee
which, if we write $\phi \equiv \rho \, \exp(i\alpha )$, 
can also be expressed as
\be
{\cal E} \equiv {\cal E}_\rho + {\cal E}_\alpha ,
\ee
where the energy density in massive modes ($\rho$) is defined as
\be
{\cal E}_\rho = 
\frac{1}{2} (\partial_t \rho )^2+\frac{1}{2} (\nabla \rho )^2 
+ \frac{\lambda}{4} (\rho^2-\eta^2)^2 ,
\label{Erho}
\ee
and that in Goldstone modes ($\alpha$) as
\be
{\cal E}_\alpha = 
\frac{\rho^2}{2} \left [ (\partial_t \alpha )^2+ (\nabla \alpha )^2 \right ] .
\label{Ealpha}
\ee

The string energy per unit length (also its tension) is found by
integrating the energy density of the solution in \eqref{stringsoln}
in the $z=0$ plane. The integration of ${\cal E}_\rho$ is finite
but the integral of ${\cal E}_\alpha$ diverges logarithmically with
distance. With a long range cutoff at $r=\Lambda$ the energy
per unit length is
\be
\mu \approx \pi \eta^2 \, \ln(\Lambda \eta) .
\label{mueq}
\ee

\begin{figure}
      \includegraphics[width=0.3\textwidth,angle=0]{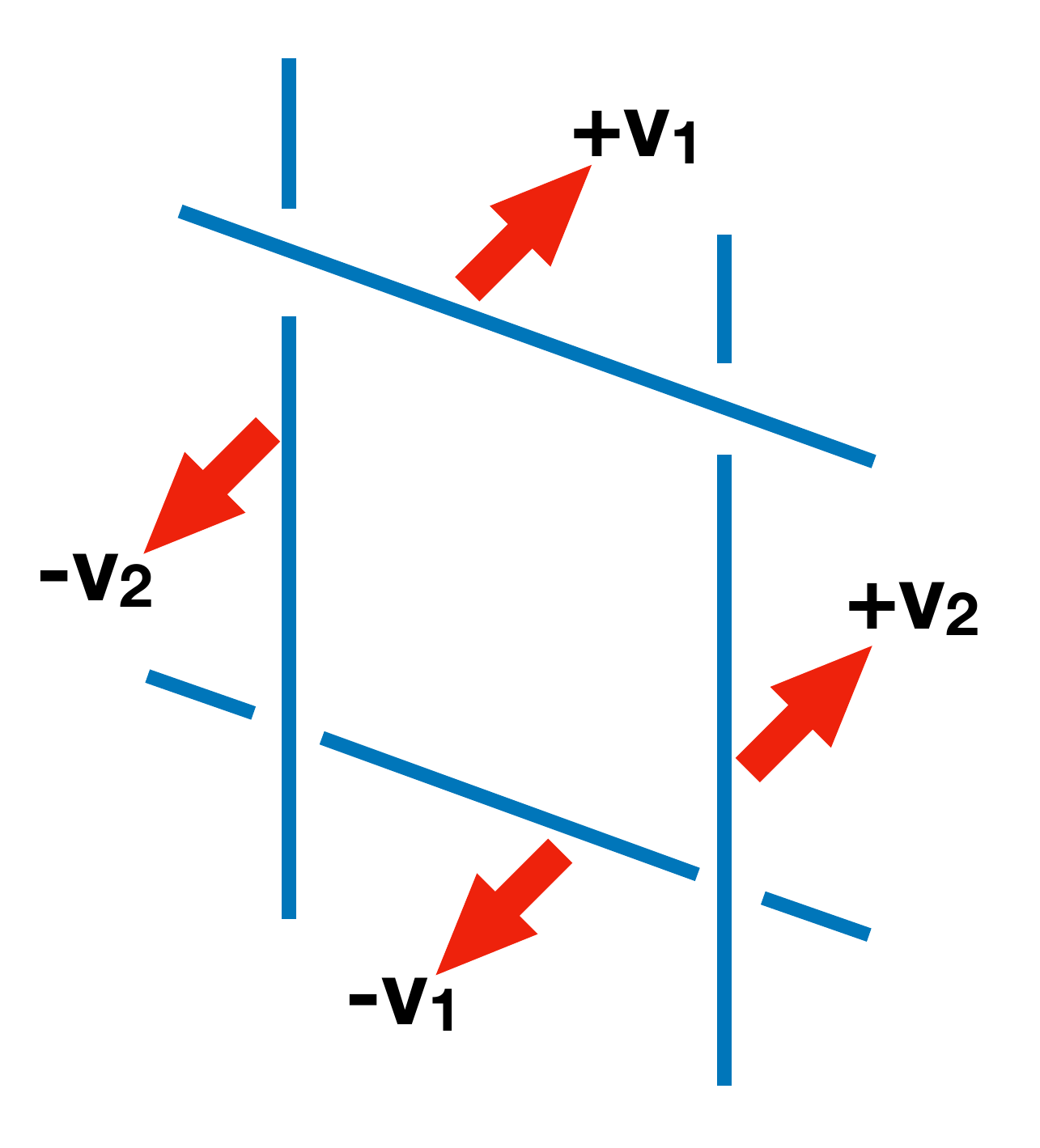}
  \caption{Four straight strings are set up with velocities as shown. 
The strings intersect and reconnect to produce a central ``inner'' loop 
and also a second ``outer'' loop because of periodic boundary conditions. 
These loops then oscillate and decay. By choosing the spacing of the
 initial straight strings, we can produce loops of different sizes, though
 they all have the same initial shape.}
\label{schematics}
\end{figure}

We now create a loop for our simulations following the scheme 
in~\cite{Matsunami:2019fss} and as illustrated in Figure~\ref{schematics}.
Our initial conditions consist of
four straight strings boosted with velocities $\pm \mathbf{v_1}$ and 
$\pm \mathbf{v_2}$ as shown schematically in Figure~\ref{schematics}. 
The four string solutions are patched together using the ``product
ansatz''. If $\Phi_a$ ($a=1,\ldots,4$) denotes the solution for the
individual strings, the field is taken to be
\be
\phi (t=0,{\bf x}) = \frac{1}{\eta^3} \prod_{a=1}^4 \Phi_a ,
\label{phi0}
\ee
and the time derivatives at the initial time are
\be
{\dot \phi} (t=0,{\bf x}) = \phi (t=0,{\bf x}) 
\sum_{b=1}^4 \frac{{\dot \Phi}_b}{\Phi_b} ,
\label{phidot0}
\ee
with ${\dot \Phi}_b$ obtained from the boosted solution
for a single string.

While this scheme can be used to construct a loop in an infinite
spatial volume, our simulations are on a finite lattice and employ 
periodic boundary conditions. These numerical limitations necessitate
some modifications of the initial conditions that are described
in Appendix~\ref{pbc}.

\section{Simulation and Results}
\label{simulationresults}

Once the four strings collide, they reconnect to form two loops -- two because 
of periodic boundary conditions. If the string velocities are 
small and not oriented suitably, the resulting loops will have insufficient 
angular momentum and will collapse quickly.
We choose velocity magnitudes that are mildly relativistic, 
$|\mathbf{v}_1|=0.6=|\mathbf{v}_2|$,
as expected in a cosmological setting.
The directions are taken to be $({\hat v}_1)_x=0.4$, 
$({\hat v}_1)_y=\sqrt{1-0.4^2} \approx 0.92$ for the two strings oriented along 
the $z$-axis and $({\hat v}_2)_z=0.4$, $({\hat v}_2)_y \approx 0.92$ for the 
strings along the $x$-axis. 

Next we use the explicit Crank-Nicholson algorithm with two iterations for the 
numerical evolution \cite{Teukolsky:1999rm} with periodic boundary conditions, 
keeping
track of the energy densities in the core of the string and the Goldstone mode
(see Eqs.~\eqref{Erho} and \eqref{Ealpha}) and the total energy and angular 
momentum. The core of the string is defined as the region where
$|\phi|/\eta < 0.9$. 
We take the initial string separation in Figure~\ref{schematics} to be half
the size of our lattice for all our runs. The simulation then produces two loops
due to the periodic boundary conditions but both loops are the same size.

 \begin{figure}
       \includegraphics[width=0.5\textwidth,angle=0]{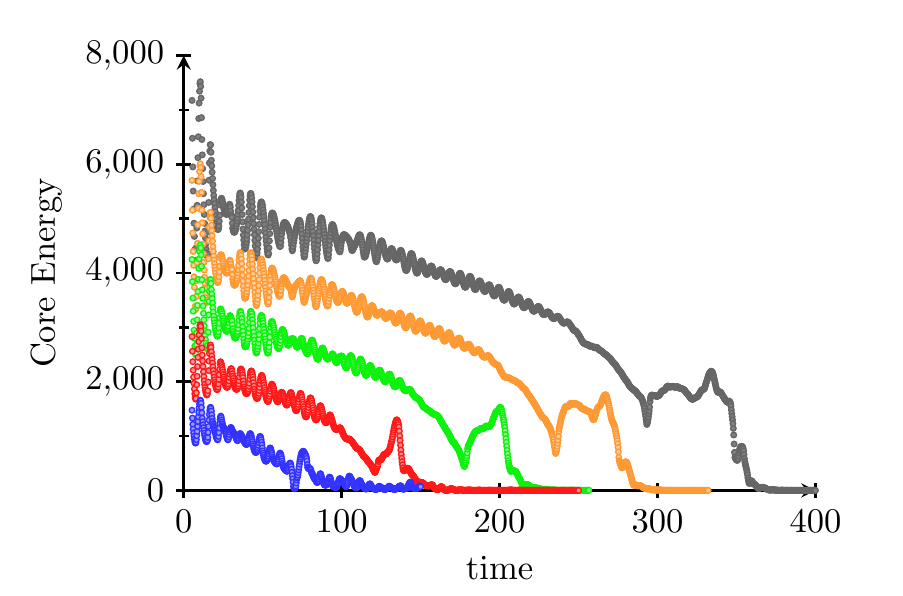}
   \caption{The loop energy as a function of time for the outer loop for 
different values of loop sizes $L=50,100,150,200,250$ (lowest to highest curve).
A similar plot is obtained for the inner loops.
	 }
 \label{coreEnergy}
\end{figure}

 \begin{figure}
       \includegraphics[width=0.5\textwidth,angle=0]{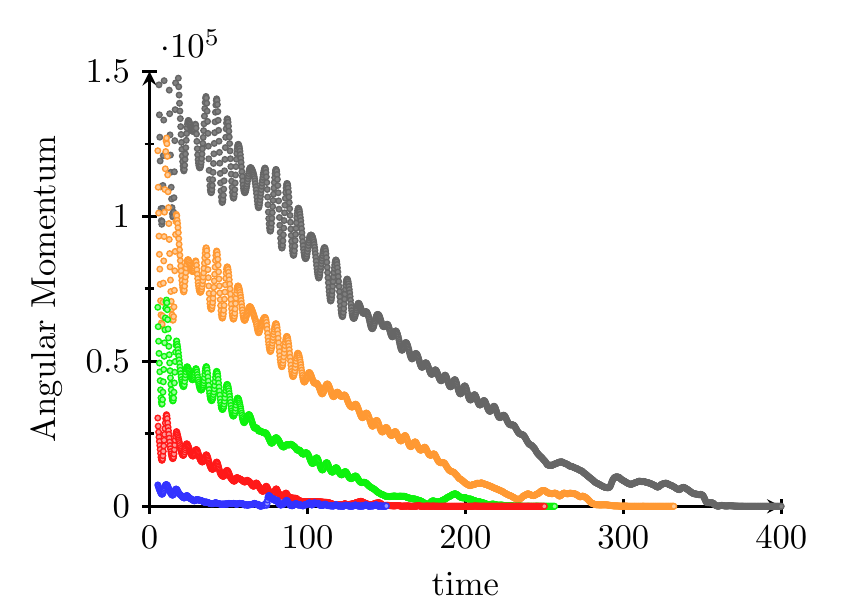}
   \caption{The loop angular momentum as a function of time for 
the outer loop for different values of loop sizes $L=50,100,150,200,250$ 
(lowest to highest curve).}
 \label{angularMomentum}
\end{figure}

We have run our simulations for a few different values of the lattice spacing, 
$\Delta x$, and found that the results are sensitive to the resolution. For example, 
the total energy in the simulation box over the entire run is conserved only at 
$\sim 20\%$ level
when $\Delta x = 0.50$ for longer runs (required for large lattices). 
We have set $\eta=1$, $e=1$, $\lambda=1/2$ and so the 
string
width is $\sim 1$. Therefore with $\Delta x=0.5$ we only have a few lattice points within the width of
the string. The run with $\Delta x=0.25$ gives better conservation, to $\sim 5\%$ level
over the entire run and agrees quite well with the 
much more computationally expensive run with $\Delta x = 0.125$.

In Figure~\ref{coreEnergy} we plot the loop energy vs time for several loop
sizes with $\Delta x=0.25$. 
We take the string core to be the lattice cells where $|\phi| < 0.9\eta$, 
with the sum of energies in all such cells giving the energy of the loop. 
Unlike in the case of gauge strings~\cite{Matsunami:2019fss}, the decay
of global string loops is not episodic and the energy gradually dissipates.

As the loops evolve, they also shed their angular momentum, defined as
\be
L_i \equiv  -\frac{1}{2} \epsilon_{ijk} \int_{\rm string\, core} \hskip -1 cm d^3x 
\, x_j \left (  \partial_t \phi \partial_k \phi^* + \partial_t \phi^* \partial_k \phi \right ).
\ee
where $x_j$ is measured from the center of energy of the loop.
In Figure~\ref{angularMomentum} we plot $|{\mathbf L}|$ versus time 
and also see gradual decay.

\section{Massive versus massless radiation}
\label{components}

The global string loop emits massive and massless Goldstone radiation.
The massive radiation corresponds to excitations of the field $\rho$ and its
energy density is given in \eqref{Erho}; the massless radiation corresponds
to excitations of $\alpha$ with energy density given in \eqref{Ealpha}. Note that $\rho$ and $\alpha$ interact, which is evident in \eqref{Ealpha}. However,
at late times, we can write $\rho = \eta + \chi$, $\theta=\alpha$, where $\chi$ is a small
excitation above the true vacuum and expand the energy density functions to
lowest order in $\chi$,
\be
{\cal E}_\rho = \frac{1}{2} \left[ (\partial_t \chi )^2+ (\nabla 
\chi )^2 +  m_\chi^2 \chi^2 \right] + \ldots \equiv {\cal E}_\chi + \ldots ,
\ee
\be
{\cal E}_\alpha = \frac{\eta^2}{2}  \left [ (\partial_t \alpha )^2+ 
(\nabla \alpha )^2 \right] + \ldots \equiv {\cal E}_\theta + \ldots ,
\ee
where $m_\chi = \sqrt{2 \lambda} \eta$.
By integrating these expressions we obtain the total energy in the two components,
\be
E_a = \int d^3x {\cal E}_a ,
\ee
where $a=\rho,\,\alpha$.
At early times, $E_\rho$ and $E_\alpha$ will differ from 
$E_\chi$ and $E_\theta$, respectively, but they will coincide at late times,
when $\rho \approx \eta$.

 \begin{figure}
       \includegraphics[width=0.45\textwidth,angle=0]{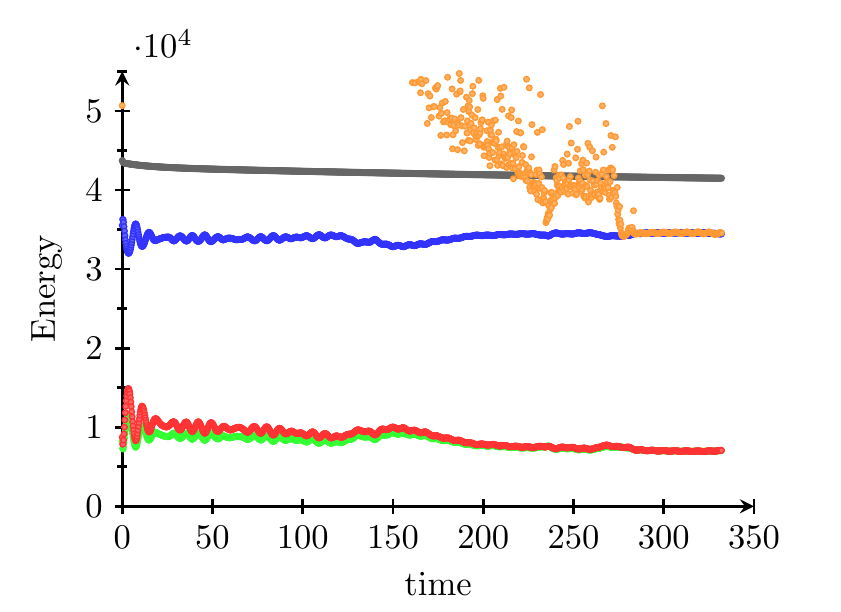}
   \caption{Energy in massive and massless components and total energy (TE) for 
   $L=200$.
   Lower curves are for massive radiation calculated
	 as $\rho$ (green) and $\chi$ (red). Middle curve is for Goldstone radiation 
	 calculated
	 for $\theta$ (blue) and $\alpha$ (orange). Top curve (black) is the total energy.}
 \label{allmodes}
\end{figure}

 \begin{figure}
       \includegraphics[width=0.45\textwidth,angle=0]{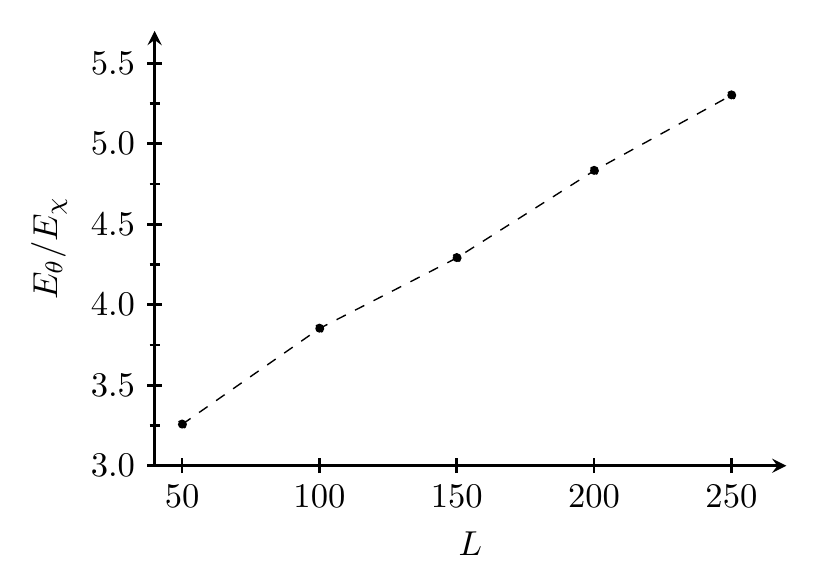}
   \caption{Plot for the ratio of energy in the Goldstone mode to 
	 energy in the massive mode at the decay time as a function of loop size.}
 \label{ratioL}
\end{figure}

 \begin{figure}
       \includegraphics[width=0.45\textwidth,angle=0]{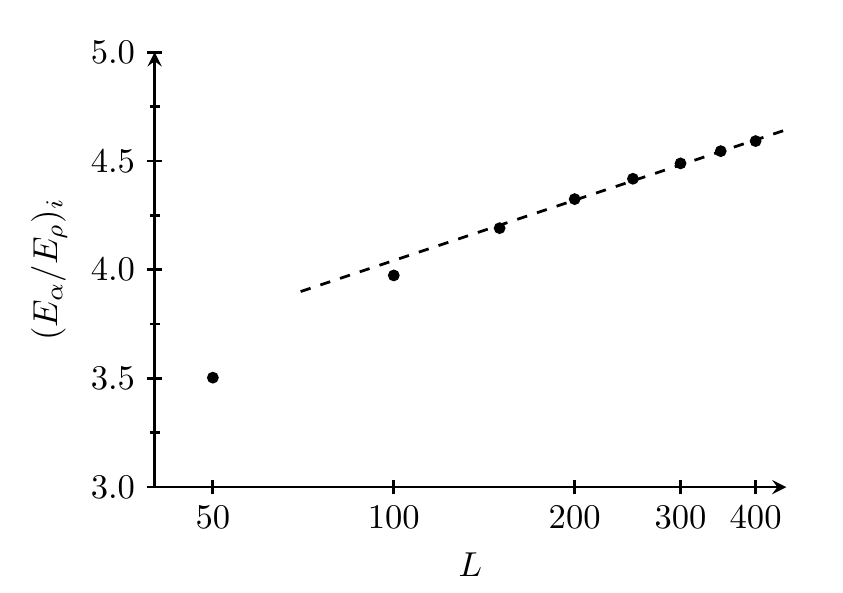}
   \caption{Log-linear plot for the ratio of initial energy in the Goldstone mode to 
	 initial energy in the massive mode as a function of loop size.}
 \label{initialratioL}
\end{figure}

In Fig.~\ref{allmodes} we plot the total energy in each of the components $\rho$, $\chi$,
$\alpha$, and $\theta$ versus time in the run with lattice size $1600^3$.
We  see that the Goldstone mode has significantly more 
initial energy compared to the massive mode and the ratio of the energies
in $\rho$ and $\theta$ remains approximately 
constant throughout the evolution.  The energies in $\chi$ and $\theta$ 
agree with those in $\rho$ and $\alpha$
once the loop has decayed as is expected.
This pattern is repeated in all our simulations with different loop sizes, however,
the ratio $E_\theta/E_\chi$ increases with loop size.
as seen in Figure.~\ref{ratioL}. This shows that massive radiation become less important
for larger loops. To obtain the length dependence of the ratio, we need to find a fit
to the plot in Figure~\ref{ratioL}. Unfortunately we could not find an unambiguous
fit to the data -- a linear dependence, power law dependence and logarithmic
dependence, all seem to fit the data equally well. Yet, based on Figure~\ref{allmodes} 
there is an alternate way to estimate the length dependence of the ratio.
This method uses the observation that the energies in $\chi$ and
$\theta$ are approximately conserved during the collapse time. 
(Eventually $\chi$ particles will decay into Goldstone bosons.)
So to obtain the ratio
we simply need to estimate these energies at the initial time, which we can
do using the initial conditions described in Sec.~\ref{model}. Since no evolution
is necessary to get the initial energies, we can go to much larger lattices. The
initial ratios versus $L$ are shown in Figure~\ref{initialratioL} on a log-linear plot,
showing that the ratio grows as $\ln(m_\chi L)$. The logarithm can also be
understood by noting that the energy of the Goldstone cloud around a single
global string diverges logarithmically with distance from the string (see \eqref{mueq}). 
For a loop, the loop size provides a cutoff on the divergence but it means that the 
Goldstone cloud has energy proportional to $\ln(m_\chi L)$. In our simulations, we 
have modified the string ansatz slightly to account for the periodic boundary conditions
as described in Appendix~\ref{pbc}, 
so we have calculated the energy numerically as shown in Figure~\ref{initialratioL}. 

 \begin{figure}
       \includegraphics[width=0.45\textwidth,angle=0]{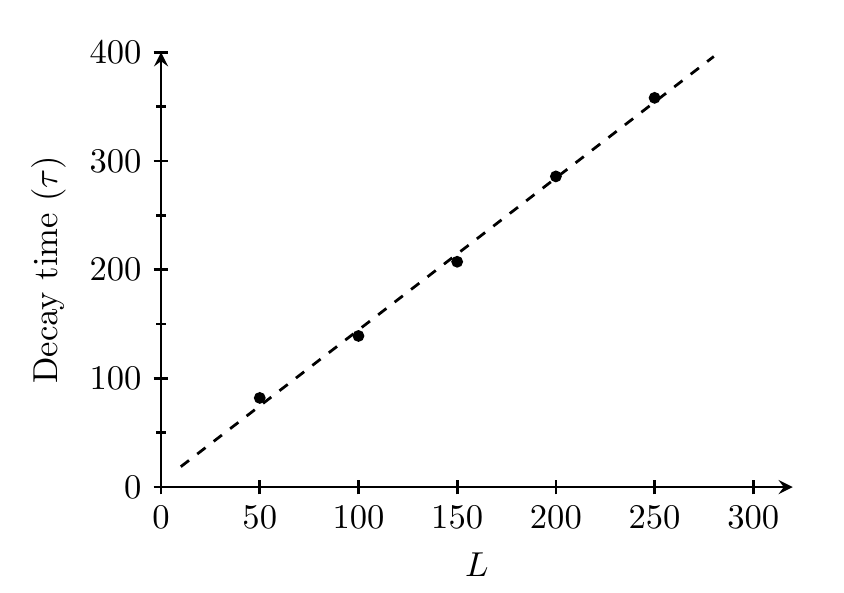}
   \caption{The loop decay time 
   as a function of the loop size.
   The best fit line is $\tau=1.39 L + 4.75$.
 }
 \label{decaytime}
\end{figure}

We use Figure~\ref{allmodes} to define the loop lifetime $\tau$: the $\theta$ and
$\alpha$ curves coincide once the strings have decayed and $\rho \approx \eta$ 
is a good approximation.
The loop lifetime $\tau$ versus loop length is shown in Figure~\ref{decaytime}
and is well-described by a linear relation $\tau \approx 1.4 L$.

\begin{figure}
\includegraphics[width=0.55\textwidth,angle=0]{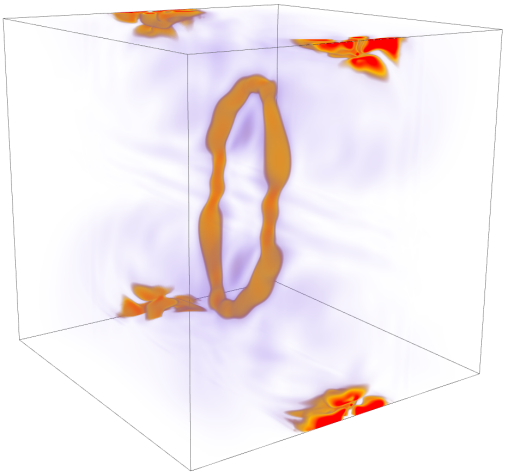}
   \caption{A snapshot of the potential energy density for a loop with $L=50$ 
	 at an intermediate time showing a ``fluffy'' deformable core and 
	 massive radiation. The full animation can be seen at~\cite{movie}.}
 \label{snapshot}
\end{figure}

In Figure~\ref{snapshot} we show a snapshot of the potential energy density at an
intermediate time in the evolution. Unlike the gauge string, the global
string is ``fluffy'', which may be understood as due to the soft power law 
profile function in \eqref{profile} as opposed to the hard exponential profile
functions in the gauge case. Deformations of the core correspond to
excitations of the massive degree of freedom. 
In the animations we see that the kinks get rounded out but they also
produce bulges in the string core as seen in Figure~\ref{snapshot}. The transfer of 
energy from kink collisions to core oscillations is an intermediate step in the process 
of the eventual decay of the entire loop energy into Goldstone modes that is not 
accounted for by the Kalb-Ramond approximation. It remains to be determined if 
the core oscillations play a significant role for cosmological size loops.

The bulges in Figure~\ref{snapshot} suggest the existence of a bound state
on a global string and we can confirm this explicitly. Consider a perturbation
of a straight global string oriented along the $z$ axis,
\be
\phi(t,r,\theta,z) = (f(r)+e^{-i\omega t} g(r)) e^{i\theta}
\ee
where we have set $\eta =1$ for convenience. 
(To obtain bound states that propagate along the string, we would replace
$\omega t$ by $\omega t - k z$.)
The string profile function satisfies,
\be
- f'' - \frac{f'}{r} + \left [ \frac{1}{r^2} + \frac{1}{2} (f^2-1) \right ] f = 0
\label{fequation}
\ee
with $f(0)=0$ and $f(\infty )=1$. Upon linearization, the perturbation $g(r)$ satisfies
the Schrodinger-type equation,
\be
-g'' - \frac{g'}{r} + \left [ \frac{1}{r^2} + \frac{3}{2} (f^2 -1) \right ] g = \Omega g
\label{gequation}
\ee
where $\Omega \equiv \omega^2 -1$ and $g(0)=0=g(\infty)$. 
A non-trivial bound state solution, {\it i.e.} with $\Omega  < 0$, 
of this equation corresponds to a bound state deformation of the global
string profile. That a bound state should exist can be seen by comparing 
Eqs.~\ref{fequation} and~\ref{gequation}. The potential term in square
brackets in the
Schrodinger equation \eqref{gequation} is deeper than the corresponding term 
appearing in
\eqref{fequation} by the extra term $2(f^2-1)/2 < 0$. 
We know that $f(\infty)=1$, so the extra term in the potential in 
\eqref{gequation} will have the effect of decreasing $g$ as compared to
$f$ and can make it vanish asymptotically for the correct eigenvalue $\Omega$.
We have confirmed this by solving
\eqref{fequation} and \eqref{gequation} and determine the lowest energy
eigenvalue to be $\Omega \approx -0.19$, thus explicitly showing the existence
of massive bound states on the global string.

\section{Energy Spectrum}
\label{spectrum}

We begin by decomposing the fields $\chi$ and $\theta=\alpha$ into Fourier modes,
\be
\chi = \int \frac{d^3k}{(2 \pi)^3} \left [ \chi_\bfk(t) e^{-i\bfk\cdot \bfx} + 
\chi_\bfk^* (t) e^{+i\bfk\cdot \bfx}  \right ]
\ee
\be
\theta = \int \frac{d^3k}{(2 \pi)^3} \left [ \theta_\bfk(t) e^{-i\bfk\cdot 
\bfx} + \theta_\bfk^* (t) e^{+i\bfk\cdot \bfx}  \right ]
\ee

The energy densities in a Fourier mode labeled by $\bfk$ are given by,
\be
{\cal E}_{\chi \bfk} =  \frac{1}{2}  \left[  |\partial_t \chi_\bfk |^2 
	+ (k^2+ m_\chi^2) | \chi_\bfk |^2 \right]
\ee
\be	
{\cal E}_{\theta \bfk} = \frac{\eta^2}{2} [ (\partial_t \theta_\bfk)^2 
	+ k^2 | \theta_\bfk |^2 ] .
\ee
In general, the spectra will depend on all
three components of the wavevector $\bfk$. However, if we sum over a
large number of loops, with different shapes, sizes and orientations, we can
expect an isotropic spectrum. To extract an isotropic spectrum from our
simulation we bin the
spectral components according to their $k=|\bfk |$ value and sum over
all modes with $|\bfk |$ in the interval ${\cal R}(k) = (k-\Delta k,k)$,
where $\Delta k = 2 \pi/(2L)$ ($2L$ is the lattice size in our 
simulations):
\be
{\cal E}_{\chi k}  = \frac{(\Delta k)^3}{(2\pi)^3} \sum_{|\bfk | \in {\cal R}(k)} {\cal E}_{\chi \bfk}, \ \ 
{\cal E}_{\theta k}  = \frac{(\Delta k)^3}{(2\pi)^3} \sum_{|\bfk |\in {\cal R}(k)}  {\cal E}_{\theta \bfk}  .
\label{calEs}
\ee
Note that the sum is over vectors $\bfk$ with the same magnitude. Hence,
it includes the $4\pi k^2$ factor that arises from the phase space volume factor
and to obtain the total energy one only needs to
sum over all the modes,
\be
E_\chi =\sum_k {\cal E}_{\chi k}, \ \ \ 
E_\theta =\sum_k {\cal E}_{\theta k}.
\ee

\begin{figure}
       \includegraphics[width=0.5\textwidth,angle=0]{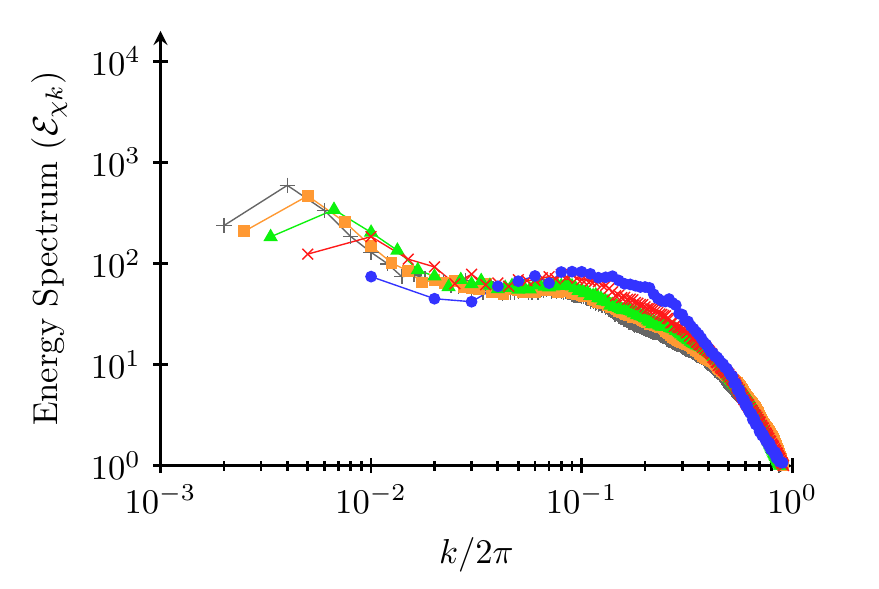}
   \caption{A log-log plot of the energy spectrum of massive radiation after the loops 
   in the simulation have collapsed for the runs with initial loop size
   50 (blue circles), 100 (red crosses), 150 (green triangles), 200 (orange 
	squares), and 250 (black pluses).  
 }
 \label{chiloglogkto1}
\end{figure}

We plot ${\cal E}_{\chi k}$ versus $k$ on a log-log scale
in Figure~\ref{chiloglogkto1}. 
The energy in the higher $k$ modes does not depend on the size of the
loop but the energy in the lowest few modes grows with the size of the loop. 
It is worth noting that the spectrum
gets cut off at $k \approx 0.1\times 2\pi \approx 0.6$ which corresponds to a 
momentum less than $m_\chi =1$.
Hence the massive particles emitted by the string are non-relativistic
especially for long loops.

 \begin{figure}
       \includegraphics[width=0.5\textwidth,angle=0]{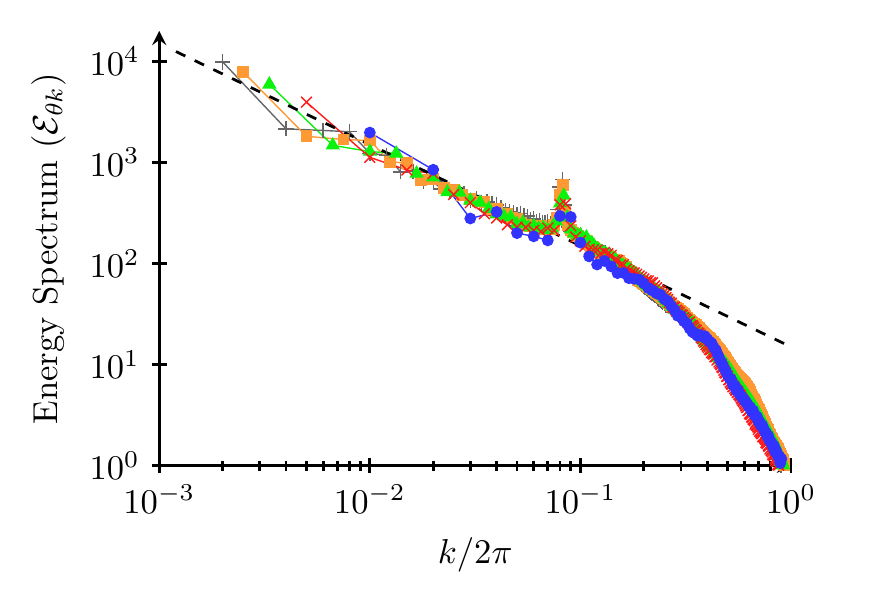}
   \caption{A log-log plot of the energy spectrum of Goldstone radiation when
   the initial loop size is 
   50 (blue circles), 100 (red crosses), 150 (green triangles), 200 (orange 
	 squares), and 250 (black pluses).  The overlaid black dashed line is 
	 given by $15/k$ and is a good fit out to $k \approx 0.08\times 2\pi 
	 \approx 0.5$. The peak at
   $k \approx 0.5$ corresponds to energy $\approx m_\chi/2$. 
   }
 \label{alphaloglogkto1}
\end{figure}

The spectrum for Goldstone radiation is shown in 
the log-log plot of Figure~\ref{alphaloglogkto1}.
We see that the spectrum decays as $15/k$ until a cutoff 
wavenumber $k_c \approx 0.5$, after which there is essentially negligible
energy contribution.
To obtain the continuum version of the energy spectrum as given in \eqref{En},
we divide both sides of \eqref{calEs} by $\Delta k = \pi/L$. Then,
\be
\frac{d{\cal E}_{\theta k}}{dk} = \frac{L}{\pi} \frac{15 \eta^2}{k}
\approx 4.8 \frac{\eta^2 L}{k}, \ \ (2L)^{-1} \le k \lesssim  m_\chi
\ee

One additional feature we see in the spectrum of Goldstone radiation is the peak at 
$k \approx 0.08\times 2\pi$ in Figure~\ref{alphaloglogkto1} for all the loops we have
simulated. The location of the peak, at $k \approx 0.5 = m_\chi/2$, reveals the origin 
of this feature. It is caused by the massive particles that decay into Goldstone bosons 
due to the interaction $\chi (\partial_\mu \theta)^2$. Since a radiated $\chi$ is 
non-relativistic, it would decay into two Goldstone bosons, each with energy
of about $m_\chi/2$.

\section{Conclusions}
\label{conclusions}

In this paper we have focused on cosmic global string loops, their dynamics
and decay. We numerically evolved loops of global string with
length up to 1000 times the width of the string core. By extrapolating
our results, we can meaningfully discuss cosmologically relevant loops
whose size can be comparable to the cosmic horizon and many orders
of magnitude larger than the core thickness.

Our results show that global string loops decay very quickly with
lifetime $\sim L$ by radiating Goldstone bosons ($\theta$) and massive
particles ($\chi$). Most of the energy is radiated in Goldstone bosons and
the emission of $\chi$ is suppressed by $1/\ln(m_\chi L)$, which in a cosmological
setting would be $\sim 0.01-0.1$. The emitted $\chi$ particles are non-relativistic
and they eventually decay into Goldstone bosons, producing a sharp peak in the
Goldstone boson energy spectrum at $m_\chi/2$. In cosmology, the decay takes 
place continually during the evolution of the global string network and the sharp 
peak at $m_\chi/2$ gets smeared out due to cosmological redshifting of the energy. 

There is one notable exception to the smearing.
If the global string network were to decay at a specific epoch, as in the axion
scenario where the network suddenly decays at the QCD epoch, then a sharp 
feature in the light particle spectrum that is produced due to the 
heavy particle decay may well survive until the present epoch. This feature
would be at high energies in the form of relativistic axions and would
be related to the Peccei-Quinn energy scales. If we denote the
Peccei-Quinn scale by $f_{\rm PQ}$, at the present epoch we would 
expect $1-10\%$ (corresponding to $1/\ln(m_\chi L)$) of the axion density to be 
in axions with energy $\sim f_{\rm PQ} /z_{\rm QCD}$ where the QCD
redshift is $z_{\rm QCD} \sim 10^{12}$. For example, with
$f_{\rm PQ} \sim10^{12}~{\rm GeV}$, the axion energy would be 
$\sim 1~{\rm GeV}$ and if axions are all the dark matter, then 1-10\% of
the dark matter would be in the form of high energy axions. We plan to 
investigate if such a spectral feature could 
be relevant to axion cosmology and axion searches in future work.

We have also extracted the energy spectrum for the Goldstone boson
radiation from global string loops in our simulations. We obtain a $1/k$ 
spectrum, confirming that the results of Refs.~\cite{Hagmann:1990mj,Hagmann:2000ja} 
hold even for loops formed by processes expected in a cosmological setting. The peak
in our simulations, at $k \approx m_\chi/2$, is a new feature, though one that should 
have been expected in hindsight. The suppressed strength of the peak (1-10\%) is 
due to the logarithmic divergence associated with the Goldstone component of
global strings.

To summarize, we have obtained a remarkably simple picture for the decay
of cosmic global string loops. Initially the loop consists of a massive field
and a Goldstone field. The loop quickly collapses, releasing radiation of
massive field and Goldstone bosons,
with the Goldstone radiation retaining its original $1/k$ spectrum. 
The radiated total energies correspond to the initial energies in these
components, except that the massive particles decay into Goldstone bosons.
As the massive particles are non-relativistic they produce a line signature in
the spectrum of Goldstone bosons at $k\approx m_\chi/2$ on top of the $1/k$
continuum. 

In contrast, studies of loop decay~\cite{Vilenkin:1986ku,Battye:1993jv}
using the Kalb-Ramond description 
assume that there is negligible massive radiation and that radiation backreaction 
can be ignored. Then it is found that all the massless radiation 
is in the first few radiation harmonics. This is at odds with our results obtained
by evolving the full field theory for loops that have length about 1000 times the
width of the string. It is clear that radiation backreaction is very
important for us, and we see a conversion of the kink collision energy into core 
oscillations that enhances massive radiation. 
Even if backreaction is included in the Kalb-Ramond approach, it does not allow 
for core oscillations and cannot account for the massive radiation. The caveat in 
the simple picture 
we have suggested is if our loops are large enough that the results can be 
extrapolated to vastly larger loops of cosmological interest.

\acknowledgements
We gratefully acknowledge Daiju Matsunami's contributions to our earlier 
collaboration \cite{Matsunami:2019fss} and thank Guy Moore, Ken Olum,
Pierre Sikivie and Alex Vilenkin for comments. AS and TV are supported 
by the U.S.  Department of Energy, Office of High Energy Physics, under 
Award No.  DE-SC0018330 at Arizona State University. LP is supported in 
part by the National Sciences and Engineering Research Council (NSERC) 
of Canada. The bulk of computations were performed on the Agave and 
Stampede2 clusters at Arizona State University and The University of Texas 
at Austin respectively.

\appendix

\section{Implementing Periodic Boundary Conditions}
\label{pbc}

The ansatz in Equations~\eqref{phi0} and \eqref{phidot0} does not satisfy
periodic boundary conditions and requires further adjustments. This is
done in several steps that we now outline. The basic idea is to first ensure
that a parallel string-antistring pair satisfies periodic boundary conditions
and then to patch together the vertical and horizontal pairs by using
the product ansatz.

We set the phase of the field and time derivatives of the field to zero at the 
boundaries of the planes orthogonal to each string-antistring pair. For 
example, for the pair along $z$ direction, we have the field
\begin{align}
	\phi_{s \bar{s},z} &= \frac{\phi_{s,z} \phi_{\bar{s},z}}{\eta} =
	\frac{|\phi_{s,z}| |\phi_{\bar{s},z}|}{\eta} e^{i (\theta_{s,z} -
	\theta_{\bar{s},z})}\\ &= \phi_{s \bar{s}, z1} +i \phi_{s \bar{s},
	z2}\nonumber
\end{align}
We now multiply the imaginary part of the field above with the function
\begin{equation}
 f_z = (1- \frac{x^2}{L^2}) (1- \frac{y^2}{L^2})
\end{equation}
where $2L$ is the size of the numerical domain. This forces the phase of
the field to vanish at the boundaries but also alters the magnitude of
the field. We then normalize the field so that the magnitude is reset to
its original value on the boundaries,
\begin{align}
	\phi_{s \bar{s}, z1}^{(1)} &= \sqrt{\frac {\phi_{s \bar{s}, z1}^2 + 
	\phi_{s \bar{s},z2}^2}{\phi_{s \bar{s}, z1}^2 + \phi_{s \bar{s},z2}^2 
	\,\, f_z^2} } \hspace{2mm} \phi_{s \bar{s}, z1} \\
	\phi_{s \bar{s}, z2}^{(1)}&= \sqrt{\frac {\phi_{s \bar{s}, z1}^2 + 
	\phi_{s \bar{s},z2}^2}{\phi_{s \bar{s}, z1}^2 + \phi_{s \bar{s},z2}^2 
	\,\, f_z^2} } \hspace{2mm} \phi_{s \bar{s}, z2} \,\, f_z \\
\end{align}
Note that the locations of the strings do not change by the above
manipulations because the zeros of the field are unaffected.

The time derivatives can now be set to zero at the boundaries by simply 
multiplying the derivatives with the function $f_z$, that is,
\begin{equation}
	\dot{\phi}_{s \bar{s}, z}^{(1)} =  \partial_t \phi_{s \bar{s}, z}^{(1)} 
	f_z
\end{equation}

At this stage the field denoted by $\phi^{(1)}_{s{\bar s},z}$ has zero phase
on the boundaries but its magnitude still varies and does not satisfy periodic
boundary conditions. In a cosmological setting the field magnitude will be
$\eta$ (=1) far from the strings. Hence we force the magnitude to be $\eta$ 
on the boundaries of the lattice. 
To achieve this, we successively divide the field $\phi^{(1)}_{s{\bar s},z}$ by 
functions that interpolate between boundary values along orthogonal directions.  
The first interpolating functions is defined as,
\be
I_{zx} = \frac{|\phi_{s \bar{s},z}^{(1)}|_{x = L} - |\phi_{s \bar{s},z}^{(1)}|_{x = -L}}{2 L} (x + L) 
+|\phi_{s \bar{s},z}^{(1)}|_{x = -L}
\ee
which gives us a new field periodic in the $x$ direction,
\begin{equation}
	\phi_{s \bar{s}, z}^{(2)}= \frac{\phi_{s \bar{s}, z}^{(1)}}{I_{zx}}.
\end{equation}
Then we divide by a corresponding interpolation function in the $y$ direction,
\be
I_{zy} = \frac{|\phi_{s \bar{s},z}^{(2)}|_{y = L} - |\phi_{s\bar{s},z}^{(2)}|_{y = -L}}{2 L} (y + L) 
+|\phi_{s \bar{s},z}^{(2)}|_{y = -L}
\ee
to get a field that satisfies periodic boundary conditions in both $x$ and $y$
directions,
\begin{equation}
	\phi_{s \bar{s}, z}^P= \frac{\phi_{s \bar{s}, z}^{(2)}}{I_{zy}}.
\end{equation}

The steps above ensure that the complex scalar field and its time 
derivatives are periodic along each direction for both string-antistring pairs.  
As a final step, we construct the field and its time derivative  using the product
ansatz,
\begin{align}
	\phi &= \frac{\phi_{s \bar{s}, z}^P \phi_{s \bar{s}, x}^P}{\eta}  \\
	\dot{\phi} &= \frac{\dot{\phi}_{s \bar{s}, z}^P \phi_{s \bar{s},
	x}^P+ \phi_{s \bar{s},
	z}^P \dot{\phi}_{s \bar{s}, x}^P}{\eta}
\end{align}
Now $\phi$ and $\dot\phi$ satisfy periodic boundary conditions and can be
used as initial conditions in our simulations.

\bibstyle{aps}
\bibliography{paper}

\end{document}